\documentclass[11pt,a4paper]{article}
\usepackage[margin=1in]{geometry}
\usepackage{amsmath,amssymb,amsfonts,bm}
\usepackage{graphicx}
\usepackage{array,multirow}
\usepackage{hyperref}
\usepackage[capitalise,nameinlink]{cleveref}
\usepackage{microtype}
\usepackage[numbers,sort&compress]{natbib}
\usepackage{caption}
\usepackage{xcolor}
\usepackage{enumitem}
\usepackage{mdframed}
\hypersetup{colorlinks=true,linkcolor=blue!50!black,citecolor=blue!50!black,urlcolor=blue!50!black}

\newcommand{\Lcal}{\mathcal{L}}

\newcommand{\dd}{\mathrm{d}}
\newcommand{\half}{\tfrac{1}{2}}
\newcommand{\Tr}{\operatorname{Tr}}
\newcommand{\pt}{\partial_t}
\newcommand{\kt}{k\partial_k}

\title{Closed-form thermal threshold functions for the \\
proper-time renormalisation group}

\author{Daniele Rizzo\thanks{\texttt{daniele.rizzo@kbfi.ee}}\\[4pt]
\normalsize\itshape Laboratory for High Energy and Computational Physics,\\
\normalsize\itshape NICPB, R\"avala 10, Tallinn 10143, Estonia}
\date{\today}

\begin{document}
\maketitle

\begin{abstract}
We derive closed-form expressions for the thermal threshold functions of the
finite-temperature proper-time renormalisation group (PTRG): until now these have been
evaluated numerically, Matsubara mode by Matsubara mode.  For the standard one-parameter
regulator family, Poisson resummation of the Matsubara sum yields a rapidly convergent
winding-number series of modified Bessel functions, and a single algebraic identity
reduces every higher threshold function to
this same closed form at a shifted kernel parameter.  The sharp proper-time regulator,
recovered as the exact $m\to\infty$ endpoint of the family with a controlled $O(1/m)$
approach, factorises into a field-dependent and a purely thermal piece.

Built on them, the local potential approximation (LPA) and its refinement to a running
anomalous dimension (LPA$'$) for the $O(N)$-symmetric theory
reduce to established results, all cross-checked against the exact Wetterich equation with
the optimised regulator.  Two findings go beyond reduction. First, the known zero-temperature
anomalous-dimension construction is extended here to finite temperature. Second, because
the closed form holds for any real regulator parameter, the refined truncation's fixed
point can for the first time be tracked as a continuous function of the regulator, rather
than at a handful of isolated points, and is found to vary smoothly and remain bounded,
with no special or pathological point anywhere on the line.
\end{abstract}

\clearpage

\section{Introduction}
\label{sec:intro}

The finite-temperature effective potential of the scalar sector controls the
thermodynamics of cosmological phase transitions, with consequences ranging from
electroweak baryogenesis~\citep{Kuzmin:1985mm,Morrissey:2012db} to the stochastic
gravitational-wave backgrounds such transitions could
source~\citep{Caprini:2019egz}.  Perturbation theory is least reliable exactly where these
questions are decided: the high-temperature loop expansion is governed by $g^2T/M$, with
$M$ the thermal mass of the light bosonic modes, rather
than by the coupling $g$ alone, infrared bosonic modes require daisy
resummation~\citep{Arnold:1992rz}, and the perturbative series for the quantities of
interest converges poorly near the transition.  Lattice simulations settle such questions
for specific models~\citep{Kajantie:1996mn} but are expensive to repeat across the parameter
spaces of extended scalar sectors, leaving non-perturbative continuum methods with
analytic control of considerable practical interest.

The functional renormalisation group (FRG) provides a possible approach.  In the formulation of
Wetterich~\citep{Wetterich:1992yh,Berges:2000ew}, the exact renormalisation group (ERG) is
built on a scale-dependent effective average
action $\Gamma_k$ that interpolates between the microscopic action at the UV scale $\Lambda$
and the full quantum effective action at $k\to0$; its flow equation is exact, and
truncations of the space of actions turn it into closed, tractable
equations.  In quantum gravity the ERG is the central tool of the
asymptotic-safety programme~\citep{Niedermaier:2006wt,Eichhorn:2026uqj,Reuter:2012id,Bonanno:2020bil}; within its truncations
the entire dynamics is encoded in a set of
\emph{threshold functions}: dimensionless loop integrals that describe how modes of
squared mass decouple from the flow at scale $k$ and temperature $T$.  With the
optimised regulator~\citep{Litim:2001up} the threshold functions are elementary at
$T=0$ and close in a single $\coth$ at finite
temperature~\citep{Berges:2000ew,Litim:2002cf}.

A closely related framework is the proper-time renormalisation group (PTRG), in which
the exact resolvent of the Wetterich equation is replaced by a Schwinger proper-time
representation of the one-loop trace~\citep{Oleszczuk:1994st,Liao:1994fp}.  The PTRG is not an
exact flow: beyond one loop the reconstructed effective action differs from the ERG one
by terms that no choice of kernel removes~\citep{Litim:2002xm,Litim:2001ky};
nevertheless, in its Wilsonian formulation the PTRG generates two-loop 
beta functions and an anomalous dimension for the $O(N)$ scalar theory 
that have been shown to be exactly correct~\citep{Giacometti:2025qyy}. 
It is UV finite kernel by kernel, analytically simple, 
and it has produced accurate critical exponents at
$T=0$ for the Ising and $O(N)$ universality classes, in truncations ranging from a
first non-trivial order in the derivative expansion~\citep{Bonanno:2000yp,
Mazza:2001bp} to a full field-dependent wave-function
renormalisation~\citep{Litim:2010tt,Bonanno:2019ukb,
Bonanno:2004pq}, as well as realistic descriptions of the
thermal chiral transition in the quark-meson
model~\citep{Bohr:2000gp,Schaefer:2001cn}.  The PTRG is also used to compute
non-perturbative effects in gravity: proper-time flows have been applied to quantum
Einstein gravity~\citep{Bonanno:2004sy}, to the gauge and parametrisation dependence
of the flow~\citep{Bonanno:2025tfj}, to regular black
holes~\citep{Bonanno:2025dry}, to non-local
truncations~\citep{Glaviano:2024hie}, and to quantum-gravity corrections to gauge
and Yukawa couplings~\citep{Giacometti:2026zrs}.  In the finite-temperature
applications, however, the thermal threshold functions have so far been evaluated
numerically, Matsubara mode by Matsubara mode, with one known exception: at the special
value $m=5/2$ a closed $\coth$ form was obtained for a quark-meson-model
application~\citep{Schaefer:2004en}.  That $m=5/2$ is where the proper-time and ERG
thresholds coincide follows from Litim's zero-temperature analysis~\citep{Litim:2001up},
where the general match condition $m=1+d/2$ between the proper-time and optimised-ERG
thresholds is derived for a $d$-dimensional theory; its $d=3$ instance fixes the leading,
dimensionally-reduced (high-temperature) piece of the coincidence.  What we show, in
\cref{sec:thresholds:def}, is that the coincidence is in fact exact at every finite $\tau$, not only
asymptotically.  

To our knowledge, for generic $m$, no closed
form of the threshold functions for the PTRG is available. In this paper we close this gap.  
Working with an $O(N)$-symmetric scalar theory with a
generic potential, we derive the thermal threshold functions of the standard
one-parameter proper-time kernel family in closed form, as rapidly convergent
winding-number series of modified Bessel functions, with a single algebraic identity
reducing every higher threshold function to this same closed form at a shifted kernel
parameter; and we show that the sharp proper-time regulator, the $m\to\infty$ endpoint of
the family, factorises into a field-dependent and a purely thermal piece, with a
controlled $O(1/m)$ approach from the smooth family.  These two closed forms are the
central results of this paper.  Built on them, assembling the finite-temperature LPA and
LPA$'$ flows is standard technique: we do so mainly to show, case by case, that the new
thresholds correctly reduce to established results and we flag explicitly the two items that are
not simply reductions: the finite-temperature anomalous dimension itself,
and the finding that the LPA$'$ fixed point, thanks to the closed form holding for any real
regulator parameter, can be tracked for the first time as a continuous function of the
regulator rather than at a handful of isolated points.

Since the proper-time flow is an approximation, internal consistency checks carry
particular weight, and we provide them throughout in the form of limiting cases: the
known zero-temperature thresholds are recovered, the flow integrated at frozen curvature
reproduces one-loop thermal perturbation theory exactly and regulator-independently, and
at high temperature the flow reduces to its three-dimensional zero-temperature
counterpart.  Throughout, the exact Wetterich flow with the optimised regulator is
carried along as a benchmark, at the level of the threshold functions and, at LPA$'$,
through a closed-form cross-check of the quartic fixed point.

The paper is organised as follows.  \Cref{sec:ptrgflow} introduces the proper-time flow.
\Cref{sec:thresholds:def} derives the basic threshold function $\Lcal_0$ and its closed
Bessel-$K$ form, including the sharp-kernel endpoint and the comparison with the exact
Wetterich flow; \cref{sec:thresholds:higher} extends this to the higher threshold
functions $\Lcal_n$ and proves the cross-family identity relating them;
\cref{sec:limits} examines the closed form's three physical limits: zero temperature,
high temperature (dimensional reduction), and heavy-mode decoupling.  \Cref{sec:lpa}
assembles the LPA truncation and its flow, the
polynomial projections, the one-loop completeness check, and the
dimensional-reduction check; \cref{sec:lpaprime} constructs the
LPA$'$ truncation and flow, the anomalous dimension, and the reduced fixed point.
\Cref{sec:discussion} summarises.  \Cref{app:bessel} derives $\Lcal_0$ and its closed
Bessel-$K$ form.

\section{The proper-time flow and its regulator family}
\label{sec:ptrgflow}

Throughout this paper we consider an $N$-component real scalar field
$\phi=(\phi_1,\dots,\phi_N)$ with an $O(N)$-symmetric potential, in $d=4$ Euclidean
dimensions; most formulas are written for a single component ($N=1$), the general case
being recovered in \cref{sec:lpa}.  Unless stated otherwise, $d=4$ holds throughout the
paper; the sole exception, the high-temperature dimensionally-reduced $3d$ system of
\cref{sec:lpa:dimred}, is always marked with an explicit ``$3d$'' superscript or subscript
and never appears otherwise.  At temperature $T$ the Euclidean time is
compact~\citep{Matsubara:1955ws}, the bosonic Matsubara frequencies are $\omega_n=2\pi nT$,
$n\in\mathbb{Z}$, and traces
run over
\begin{equation}
  \Tr_T[\,\cdot\,] \;=\; T\sum_{n=-\infty}^{\infty}\int\!\frac{\dd^3p}{(2\pi)^3}\,
  \operatorname{tr}_{\rm field}[\,\cdot\,].
\end{equation}
Scale dependence is expressed through the RG time $t=\ln(k/\Lambda)\le0$, with
$\pt\equiv\kt$; the flow runs from the UV scale $k=\Lambda$ ($t=0$) to the IR $k\to0$
($t\to-\infty$).

The PTRG constructs a flow equation for a
Wilsonian action $S_k$, which retains an explicit reference to the microscopic scale
$\Lambda$ at which the bare action is specified, and is intended to be substituted into the
remaining path integral over modes below $k$, rather than read off directly as the
generator of one-particle-irreducible correlators~\citep{Bonanno:2019ukb}.  This is achieved by
replacing the resolvent of the ERG with a Schwinger
proper-time representation of the one-loop trace.  The PTRG flow is postulated in the
form~\citep{Oleszczuk:1994st,Liao:1994fp}
\begin{equation}
  \pt S_k \;=\; \half\,\Tr_T\int_0^\infty\frac{\dd s}{s}\,
  \bigl(\pt f_k\bigr)(s)\;e^{-s\,S_k^{(2)}}
  \label{eq:ptflow}
\end{equation}
where $f_k(s)$ is a proper-time blocking function and $S_k^{(2)}\equiv\delta^2S_k/\delta\phi\,\delta\phi$
is the second functional derivative of the Wilsonian action, evaluated on the background
field configuration.  We work directly with the
$t$-derivative of the blocking function, i.e.\ the kernel $(\pt f_k)(s)$ is the primary
object; this is the convention of the thermal PTRG
literature~\citep{Bohr:2000gp,Schaefer:2001cn} and it matters at
LPA$'$, where $Z_k$-dependent kernels would otherwise generate additional
$\eta$-proportional terms upon differentiation.

The standard one-parameter family of regulators is built from the incomplete Gamma
function, with two reference scales: the running scale $k$ and the microscopic scale
$\Lambda$ at which $S_k$ is specified,
\begin{equation}
  \rho_{k,\Lambda}(s;m) \;\equiv\; \frac{\Gamma(m,sk^2)-\Gamma(m,s\Lambda^2)}{\Gamma(m)} ,
  \qquad \Gamma(m,x)\equiv\int_0^x\dd y\,y^{m-1}e^{-y} ,
  \label{eq:wilsonianreg}
\end{equation}
which vanishes for $s>1/\Lambda^2$ (UV) and $s<1/k^2$ (IR) and interpolates smoothly in
between~\citep{Bonanno:2019ukb}.  Since $\Lambda$ is fixed and the kernel
carries no explicit $Z_k$, only the first term contributes to its $t$-derivative,
\begin{equation}
  \bigl(\pt f_k\bigr)(s;m) \;=\; k\partial_k\rho_{k,\Lambda}(s;m) \;=\;
  \frac{2}{\Gamma(m)}\,(sk^2)^m\,e^{-sk^2},
  \qquad m>\tfrac32 .
  \label{eq:mkernel}
\end{equation}
In terms of the dimensionless proper time $u=sk^2$ the kernel
$F(u;m)=2u^me^{-u}/\Gamma(m)$ is a normalised bump peaked at $u^*=m$, i.e.\ at
$s^*=m/k^2$: the flow at scale $k$ is dominated by fluctuations of covariance
$\sim k^2/m$.  The constraint $m>3/2$ makes the spatial momentum integral converge; the
stronger condition $m>2$, assumed from now on, additionally makes the $T=0$ ($n=0$
Matsubara) contribution finite.  The value $m=3$ is distinguished: in $d=4$ at $T=0$ it
reproduces, at LPA, exactly the flow of the Wetterich equation~\eqref{eq:wetterich} with
the four-dimensional Litim regulator~\citep{Litim:2002xm}.  Since~\eqref{eq:mkernel} depends only on $k$, every threshold
function, fixed point, and critical exponent derived below is independent of $\Lambda$;
equivalently, nothing changes if $S_k$ is read as a single-scale, effective-average-action-style
object with no $\Lambda$ at all, so that all results obtained here within the Wilsonian
approach can be ported directly to the effective-average-action reading with no
modification.  The literature also uses an equivalent, differently normalised member of
this same one-parameter family, in which an extra factor of $m$ rescales the proper-time
argument of~\eqref{eq:wilsonianreg}~\citep{Bonanno:2019ukb}; this amounts
to relabelling the running scale by $k\to k/\sqrt m$ and leaves every physical,
dimensionless result unaffected.

The second regulator used throughout is the \emph{sharp} proper-time kernel,
\begin{equation}
  \bigl(\pt f_k^{\rm sharp}\bigr)(s) \;=\; 2\,\delta(sk^2-1),
  \label{eq:sharpkernel}
\end{equation}
which fires at the single proper time $s=1/k^2$.  Both kernels share a normalisation
property that plays a central role below: for every fixed $s>0$,
\begin{equation}
  \int_0^\infty\frac{\dd k}{k}\;\bigl(\pt f_k\bigr)(s) \;=\; 1 ,
  \label{eq:partition}
\end{equation}
as follows by substituting $y=sk^2$, which gives
$\frac{1}{\Gamma(m)}\int_0^\infty\dd y\,y^{m-1}e^{-y}=1$ for the $m$-family and
$\int_0^\infty\frac{\dd y}{y}\,\delta(y-1)=1$ for the sharp kernel. Equation~\eqref{eq:partition} states that the kernels resolve the identity along the RG trajectory:
summing the flow over all scales at \emph{frozen} $S_k^{(2)}$ reconstructs the full
proper-time representation of the one-loop functional determinant,
$-\half\Tr\int_0^\infty\frac{\dd s}{s}e^{-sS^{(2)}} = -\half\Tr\ln S^{(2)}$ (up
to the usual UV subtractions carried by the $s\to0$ endpoint).  This is the precise sense
in which the PTRG is \emph{one-loop complete}: it reproduces one-loop perturbation theory
exactly, for every admissible kernel.  \Cref{sec:lpa} turns this observation
into a quantitative finite-temperature check.  Beyond one loop the PTRG flow
\eqref{eq:ptflow} is an approximation: it cannot be derived from an exact
renormalisation-group flow except in particular limits, and it misses specific two-loop
contributions that the Wetterich flow captures
\citep{Litim:2002xm,Litim:2001ky}.  All PTRG statements in this paper are
therefore statements about a \emph{defined approximation scheme}, whose quality must be
assessed by comparison with the exact flow and with known limits, as we do throughout.

The benchmark flow is the Wetterich equation, to which we now turn.  Unlike the
Wilsonian action $S_k$ above, the ERG
is formulated for the effective average action
$\Gamma_k[\phi]$, the generator of one-particle-irreducible correlators at scale $k$ with no residual
$\Lambda$-dependence, obtained by adding to the microscopic action an IR regulator term
$\Delta S_k=\half\int_q\phi(-q)R_k(q^2)\phi(q)$ and performing a modified Legendre
transform.  Its scale dependence is governed by the Wetterich
equation~\citep{Wetterich:1992yh}
\begin{equation}
  \pt \Gamma_k \;=\; \half\, \Tr_T
  \Bigl[\bigl(\Gamma_k^{(2)} + R_k\bigr)^{-1}\pt R_k\Bigr],
  \qquad \Gamma_k^{(2)}\equiv\delta^2\Gamma_k/\delta\phi\,\delta\phi \, .
  \label{eq:wetterich}
\end{equation}
Equation~\eqref{eq:wetterich} is exact: the right-hand side has one-loop \emph{structure},
but the loop is built on the full, field-dependent, scale-dependent propagator
$(\Gamma_k^{(2)}+R_k)^{-1}$, so the flow resums arbitrarily high loop orders once it is
integrated.  The optimised Litim regulator~\citep{Litim:2001up},
\begin{equation}
  R_k^{\rm opt}(q^2)=(k^2-q^2)\,\Theta(k^2-q^2),
  \label{eq:litimreg}
\end{equation}
makes the regulated propagator momentum-independent inside the shell $q^2<k^2$ and is the
reference point for all ERG formulas in this paper.  At finite temperature we take
\eqref{eq:litimreg} to act on the spatial momenta only, which is the standard choice
compatible with the broken $O(4)$ invariance of the thermal state; the consequences of
this choice are discussed in \cref{sec:thresholds:def}.  The asymmetry between the
compact temporal direction and the non-compact spatial ones is physical here, since the
thermal ensemble itself breaks $O(4)$, unlike the case of compact extra dimensions,
where treating the mode sum and the momentum integral asymmetrically is known to change
the result~\citep{Branchina:2023rgi}.

\section{Thermal threshold functions of the proper-time family}
\label{sec:thresholds}

\subsection{The threshold function $\Lcal_0$: definition and closed forms}
\label{sec:thresholds:def}

Consider the local-potential ansatz for a constant background field, on the Euclidean
thermal cylinder of circumference $\beta=1/T$ in imaginary time $\tau\in[0,\beta)$,
\begin{equation}
  S_k[\phi] \;=\; \int_0^\beta\!\dd\tau\int\!\dd^3x\,
  \Bigl[\tfrac12(\partial_\mu\phi)^2+U_k(\phi)\Bigr],
  \label{eq:lpaansatz}
\end{equation}
so that $S_k^{(2)}(p_0,\mathbf p)=\mathbf p^2+\omega_n^2+U_k''(\phi)$.  Projecting
\eqref{eq:ptflow} onto this ansatz with the kernel~\eqref{eq:mkernel}
defines the basic thermal threshold function $\Lcal_0$ through
\begin{equation}
  \kt U_k(\phi;T)\;=\;k^4\,\Lcal_0(w;\tau),
  \qquad w=\frac{U_k''(\phi)}{k^2},\quad\tau=\frac{T}{k} .
  \label{eq:L0def}
\end{equation}
From here on $\tau$ denotes exclusively this dimensionless temperature $T/k$, not the
imaginary-time coordinate of~\eqref{eq:lpaansatz}, which plays no further role below.
Carrying out the proper-time and spatial integrals (the elementary steps are recalled at
the start of \cref{app:bessel}) gives the Matsubara representation
\begin{equation}
  \Lcal_0^{(m)}(w;\tau)
  = \frac{\sqrt{\pi}\,\Gamma(m-\tfrac32)}{8\pi^2\,\Gamma(m)}\;\tau
  \sum_{n=-\infty}^{\infty}
  \bigl(1+w+4\pi^2n^2\tau^2\bigr)^{3/2-m},
  \label{eq:matsubara}
\end{equation}
valid for $m>3/2$, $w>-1$ and $\tau>0$; the sum converges absolutely for $m>2$.  It is a
sum over a quadratically shifted argument, $\sum_n(a+n^2)^{p}$, whose exponent
$p=3/2-m$ is half-integer for integer $m$, unlike the ERG case, where the corresponding
exponent is $-1$ and partial fractions apply.  This is the structural reason why no
elementary closed form exists, and why the natural resummation produces Bessel functions
instead.  The Matsubara sum can nevertheless be evaluated in closed form (a
non-elementary one, in terms of Bessel functions) for $m>2$, $w>-1$ and $\tau>0$ we
find
\begin{equation}
\Lcal_0^{(m)}(w;\tau)
  = \frac{1}{16\pi^2\,\Gamma(m)}\left[
  \frac{\Gamma(m-2)}{(1+w)^{m-2}}
  \;+\;4\sum_{\ell=1}^{\infty}
  \left(\frac{\ell}{2\tau\sqrt{1+w}}\right)^{m-2}
  K_{m-2}\!\left(\frac{\ell\sqrt{1+w}}{\tau}\right)\right],
  \label{eq:besselform}
\end{equation}
which is the central result of this section.  The derivation (details are presented in \cref{app:bessel}) proceeds
by writing each Matsubara term of~\eqref{eq:matsubara} as a Mellin--Laplace integral,
recognising the Matsubara sum under the integral as a Jacobi theta function, applying the
modular (Poisson) inversion~\citep{DLMF}
\begin{equation}
  \sum_{n\in\mathbb{Z}}e^{-4\pi^2\tau^2tn^2}
  =\frac{1}{2\tau\sqrt{\pi t}}\sum_{\ell\in\mathbb{Z}}e^{-\ell^2/(4\tau^2 t)},
  \label{eq:jacobi}
\end{equation}
and evaluating the resulting integrals with a standard tabulated Bessel
identity~\citep{DLMF}.  The index
$\ell$ is the number of windings of the thermal loop around the Euclidean time circle: the
$\ell=0$ term is the zero-temperature contribution, and each winding $\ell\ge1$ is
suppressed by $e^{-\ell\sqrt{1+w}/\tau}=e^{-\ell\,M_{\rm eff}/T}\,(1+\dots)$ with
$M_{\rm eff}^2=k^2(1+w)$ --- Boltzmann suppression of thermal excitations of the regulated
mode.

The $m$-family threshold is defined only for $w>-1$, where the regulated inverse
propagator $k^2(1+w)$ stays positive, and it diverges as this edge is approached,
\begin{equation}
  \Lcal_0^{(m)}(w;\tau)\;\sim\;
  \frac{\Gamma(m-2)}{16\pi^2\Gamma(m)}\,(1+w)^{2-m}
  \;+\;
  \frac{\sqrt{\pi}\,\Gamma(m-\tfrac32)}{8\pi^2\Gamma(m)}\,
  \tau\,(1+w)^{3/2-m},
  \label{eq:wminus1}
\end{equation}
the thermal piece enhanced by half a power relative to the $T=0$ term --- the
three-dimensional singularity of the Matsubara zero mode.  The divergence is physical,
not pathological: in the non-convex region of a potential describing coexisting phases the
curvature $w=U_k''/k^2$ is driven towards $-1$, the source above grows without bound, and the
flow flattens the potential, restoring convexity dynamically as
$k\to0$~\citep{Berges:2000ew}.  Any numerical prescription for handling the approach to
$w=-1$ (for example stopping the flow at a scale $k_f$, or clamping $w$ from below or resolving the
full flattening on a grid) is itself a choice of coarse-graining scheme, and
quantities that live in the inner region (barrier heights, degeneracy points of
non-convex branches) depend on it; only outer-region quantities are scheme-independent in
this sense.  The sharp kernel, considered next, has no analogue of this singularity.

The sharp kernel~\eqref{eq:sharpkernel} yields a threshold function of quite different
analytic character.  There the proper-time integral collapses to $s=1/k^2$, the spatial
integral is Gaussian, and Poisson resummation of the remaining Matsubara sum gives
\begin{equation}
\Lcal_0^{\rm sharp}(w;\tau)
  = \frac{e^{-w}}{16\pi^2}\,G(\tau),
  \qquad
  G(\tau)=1+2\sum_{\ell=1}^{\infty}e^{-\ell^2/(4\tau^2)}
  =\vartheta_3\!\bigl(0,e^{-1/(4\tau^2)}\bigr),
  \label{eq:sharp}
\end{equation}
with $\vartheta_3(z,q)\equiv\sum_{n=-\infty}^{\infty}q^{n^2}e^{2inz}$ the Jacobi theta
function~\citep{DLMF}.  The $w$- and $\tau$-dependences factorise exactly, with a thermal factor that
interpolates between $G(0)=1$ and $G(\tau)\to2\sqrt{\pi}\,\tau$ at high temperature.  Unlike
\eqref{eq:wminus1}, this expression is entire in $w$, growing only as $e^{|w|}$ for
$w<0$. At $w=0$, $\tau=0$ the sharp threshold is $1/(16\pi^2)$, exactly twice
the $m=3$ value $1/(32\pi^2)$.
No finite-$m$ member of the family shares the sharp kernel's exact factorisation: by
\eqref{eq:wminus1}, the vacuum and thermal pieces decouple in $w$ with different powers,
$(1+w)^{2-m}$ and $(1+w)^{3/2-m}$ respectively, which no factorising ansatz $F(w)\,G(\tau)$
can reproduce for any finite $m$.

The sharp kernel is not an isolated scheme but the endpoint of the $m$-family: the threshold function~\eqref{eq:sharp} can also be obtained as
the $m\to\infty$ limit of the closed form~\eqref{eq:besselform}, provided the limit is
taken with care.  Taking $m\to\infty$ naively,
at fixed arguments $(w,\tau)$, simply gives zero, since the kernel $F(u;m)$ concentrates at
$u^*=m$ with relative width $m^{-1/2}$, i.e.\ at the proper time $s^*=m/k^2$, which
drifts away from $s=1/k^2$ as $m$ grows.  The limit must instead be taken at fixed
physical scales, letting the coarse-graining scale absorb the drift.  Defining
$\bar k^2\equiv k^2/m$, so that $s^*=1/\bar k^2$ fires at the sharp proper time of the
scale $\bar k$, one has, for fixed $U_k''$ and $T$,
\begin{equation}
  k^4\,\Lcal_0^{(m)}\!\left(\frac{U''}{k^2};\frac{T}{k}\right)
  \;\xrightarrow[m\to\infty]{}\;
  \bar k^4\,\Lcal_0^{\rm sharp}\!\left(\frac{U''}{\bar k^2};\frac{T}{\bar k}\right),
  \label{eq:sharplimit-flow}
\end{equation}
with relative corrections of order $1/m$.  At $T=0$ the limit is elementary: writing
$\bar w\equiv U''/\bar k^2$,
$m^2\,\Lcal_0^{(m)}(\bar w/m;0)=\frac{m^2}{16\pi^2(m-1)(m-2)}\,(1+\bar w/m)^{2-m}
\to e^{-\bar w}/(16\pi^2)$, precisely the sharp threshold; the finite-temperature
statement follows analogously from the uniform large-order asymptotics of $K_\nu$.

\paragraph{Comparison with the exact Wetterich flow.}
Throughout this paper we use the exact Wetterich flow as a running benchmark, comparing
to it both at the level of the threshold functions and at
the level of fixed-point exponents, where repeating a computation with the exact flow at
identical truncation separates truncation artefacts from genuine defects of the
proper-time approximation.  For the Wetterich equation~\eqref{eq:wetterich} with the
Litim regulator~
\eqref{eq:litimreg} acting on spatial momenta, the LPA projection can be summed in closed
form.  Within the shell $|\mathbf q|<k$ the regulated inverse propagator is the
momentum-independent constant $k^2+\omega_n^2+U_k''$, the spatial integral produces only
the ball volume $k^3/(6\pi^2)$, and the Matsubara sum is rational in $n^2$ and hence
partial-fraction summable via the standard sum $\sum_{n\in\mathbb Z}(a+n^2)^{-1}
=\pi\coth(\pi\sqrt a)/\sqrt a$~\citep{Berges:2000ew}.  The result is
\begin{equation}
\Lcal_0^{\rm ERG}(w;\tau)
  \;=\;\frac{\coth\!\bigl(\sqrt{1+w}/(2\tau)\bigr)}{12\pi^2\,\sqrt{1+w}}\;,\qquad
  \kt U_k = k^4\,\Lcal_0^{\rm ERG}(w;\tau),
  \label{eq:ergcoth}
\end{equation}
with the limits
\begin{equation}
  \Lcal_0^{\rm ERG}(w;0)=\frac{1}{12\pi^2\sqrt{1+w}},\qquad
  \Lcal_0^{\rm ERG}(w;\tau)\xrightarrow{\tau\to\infty}
  \tau\,\underbrace{\frac{1}{6\pi^2(1+w)}}_{\displaystyle\equiv\;\ell_0^{3d,\rm ERG}(w)}
  +\frac{1}{72\pi^2}\frac{1}{\tau}+O(\tau^{-3}).
  \label{eq:erglimits}
\end{equation}
In the notation of~\citep{Berges:2000ew} this is
the statement $\ell_0^{4,T}\to\tau\,\ell_0^{3}$ at high temperature. The coefficient $\ell_0^{3d,\rm ERG}(w)$ is the ERG's own
dimensional-reduction constant, distinct from the proper-time family's own high-temperature
coefficient $\ell_0^{3d}(w)$; the two coincide only at
$m=5/2$, as~\eqref{eq:m52ergcoth} makes exact.

The structural contrast with~\eqref{eq:besselform} is clean.  In the ERG the regulated
propagator is \emph{rational} in the Matsubara frequency, the thermal sum is
$\sum_n(a+n^2)^{-1}=\pi\coth(\pi\sqrt a)/\sqrt a$, and the threshold closes in a single
elementary function.  In the PTRG the mode suppression is \emph{Gaussian} in proper time,
each Matsubara mode contributes a half-integer power $(A_n)^{3/2-m}$, and the closure
requires the full modular machinery~\eqref{eq:jacobi}, producing Bessel functions of the
winding number.  The two representations organise the same physics in dual ways: the
$\coth$ form is a statement in frequency space (all Matsubara modes summed), the Bessel
form a statement in winding-number space (Boltzmann factors of multiply wound thermal
trajectories).  Each is exponentially convergent precisely where the other is slowly
convergent.

At zero temperature the two frameworks can be made to coincide.  With the full
$O(4)$-invariant Litim regulator, the ERG LPA flow is
$\kt U_k=k^4/\bigl(32\pi^2(1+w)\bigr)$, identical to the $T=0$ limit of the
$m=3$ closed form~\eqref{eq:besselform}, $\Lcal_0^{(3)}(w;0)=1/(32\pi^2(1+w))$: at this
level the two schemes are indistinguishable, which is the LPA
shadow of the general map between proper-time flows and background-field ERG flows
constructed in~\citep{Litim:2002xm}.  The same coincidence is observed directly for the same
one-parameter proper-time kernel family used here, at the level of the Wilsonian
flow equation itself~\citep{Bonanno:2019ukb}.  At finite temperature the coincidence at
$m=3$ is broken twice over.  First, the thermal ERG regulator acts on spatial momenta
only, so already its $T=0$ limit, $1/(12\pi^2\sqrt{1+w})$ in~\eqref{eq:erglimits},
differs from the $O(4)$ form.  Second, for generic $m$ the $m$-family threshold is not, unlike
\eqref{eq:ergcoth}, a function of the single combination $\sqrt{1+w}/\tau$ times
$(1+w)^{-1/2}$, so the
$m=3$ coincidence does not persist at $T>0$.

\begin{figure}[t]
\centering
\includegraphics[width=0.62\textwidth]{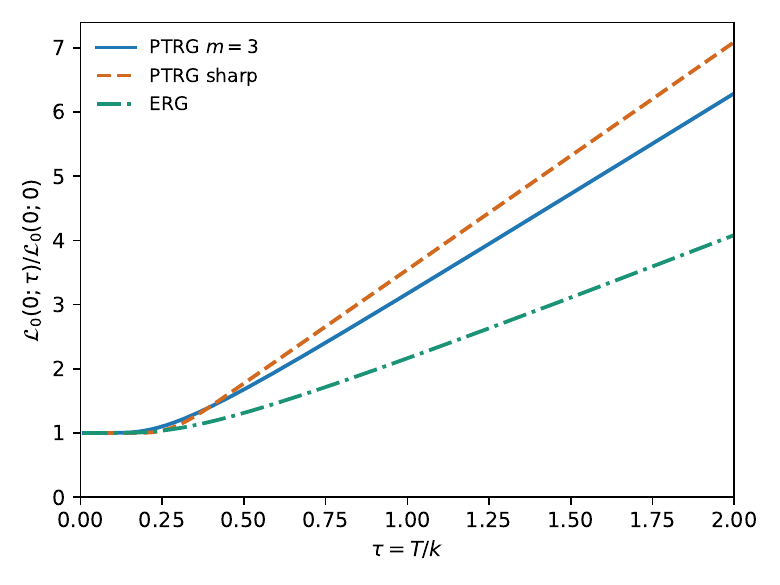}
\caption{Thermal threshold functions at vanishing curvature, normalised to their
$T=0$ values, $1/(32\pi^2)$ ($m=3$, solid blue), $1/(16\pi^2)$ (sharp, dashed
orange) and $1/(12\pi^2)$ (ERG, dot-dashed green).  ``ERG'' denotes
the exact Wetterich flow with the optimised Litim regulator acting on spatial momenta,
i.e.~\eqref{eq:ergcoth}.
}
\label{fig:L0ratio}
\end{figure}

There is, however, exactly one point of the regulator line where the coincidence with the
ERG is restored, and there it holds identically in $w$ and $\tau$, not just
asymptotically.  At $m=5/2$ the Matsubara exponent $3/2-m$ in~\eqref{eq:matsubara} equals
$-1$, precisely as for the ERG; the Bessel index in
\eqref{eq:besselform} is then $m-2=\tfrac12$, where $K_{1/2}(z)=\sqrt{\pi/(2z)}\,e^{-z}$ is
elementary, and the winding sum collapses to a geometric series.  Carrying this through
gives
\begin{equation}
  \Lcal_0^{(5/2)}(w;\tau)=\Lcal_0^{\rm ERG}(w;\tau)
  \label{eq:m52ergcoth}
\end{equation}
identically; in particular, taking $\tau\to\infty$ on both sides recovers
$\ell_0^{3d}(w)\big|_{m=5/2}=\ell_0^{3d,\rm ERG}(w)$, with the ERG coefficient
defined in~\eqref{eq:erglimits}.  The leading, $\tau\to\infty$ piece of this coincidence
was already implicit in~\citep{Litim:2001up}, where the general match condition
$m=1+d/2$ between the proper-time and optimised-ERG thresholds at $T=0$ is derived for a
$d$-dimensional theory; what is new here is that the coincidence
persists exactly at every finite $\tau$, not only asymptotically.  This same exact closed
form was also obtained, independently and by direct
Matsubara summation rather than through~\eqref{eq:besselform},
for a quark-meson-model application
in~\citep{Schaefer:2004en}: the optimised blocking function used there, chosen so as to
make the finite-$T$ flow analytic, is precisely the $m=5/2$ member of the kernel
family~\eqref{eq:mkernel}, and its finite-temperature pion/sigma threshold 
structure is the $O(4)$ instance of the generic $O(N)$ LPA flow, evaluated with
the $m=5/2$ threshold, which by~\eqref{eq:m52ergcoth} coincides exactly with the
ERG one.  Away from this
single point of the regulator line, the finite-temperature PTRG is a genuinely different
regularisation scheme, not a rewriting of the thermal ERG.

The quantitative content of the comparison is collected in \cref{fig:L0ratio}.  The raw
$T=0$ values, $\Lcal_0(0;0)=1/(32\pi^2)$, $1/(16\pi^2)$ and $1/(12\pi^2)$ for $m=3$,
sharp and ERG respectively, do not coincide, and the raw ordering is not even
temperature-independent: the ERG is the largest threshold at low $\tau$ but is overtaken
by the sharp PTRG near $\tau\approx0.5$, in line with the high-temperature slopes
$\tau/(8\pi^{3/2})>\tau/(6\pi^2)>\tau/(32\pi)$ implied by the respective
dimensional-reduction limits.  These absolute normalisations are, however, pure scheme
redefinitions of $k$; physical statements must, and in the flows below do, compare
dimensionless IR observables rather than raw threshold magnitudes.
\Cref{fig:L0ratio} therefore shows each threshold divided by its own $T=0$ value.  Once
the scheme normalisations are factored out, the two proper-time schemes nearly coincide
over the whole temperature range (their asymptotic slopes, $2\sqrt{\pi}$ for
sharp and $\pi$ for $m=3$, differ by only $13\%$), all three curves share the
Boltzmann-flat plateau below $\tau\lesssim1/(2\pi)$, and the genuine outlier in thermal
response is the ERG with its markedly weaker slope $2$.  Thermal effects dominate the
flow well before the naive $T=k$ crossing and most of the apparent spread
among the raw thresholds is thus normalisation convention; the residual, physical
difference is between the proper-time family as a whole and the Wetterich flow.

\subsection{Higher threshold functions and the cross-family identity}
\label{sec:thresholds:higher}

Polynomial projections of the flow and the LPA$'$ anomalous dimension require the
$w$-derivatives of $\Lcal_0$.  Define
\begin{equation}
  \Lcal_n(w;\tau)\;\equiv\;\frac{(-1)^n}{n!}\,
  \frac{\partial^n\Lcal_0}{\partial w^n}(w;\tau),\qquad n\ge0 .
  \label{eq:Lndef}
\end{equation}
With this convention all $\Lcal_n$ are positive and the beta
functions below carry no stray signs.  Iterating the Bessel recursion
$\dd[z^{-\nu}K_\nu(z)]/\dd z=-z^{-\nu}K_{\nu+1}(z)$~\citep{DLMF} through~\eqref{eq:besselform} gives
the closed form
\begin{equation}
  \Lcal_n^{(m)}(w;\tau)
  = \frac{1}{16\pi^2\,\Gamma(m)}\left[
  \frac{\Gamma(m+n-2)}{n!\,(1+w)^{m+n-2}}
  +\frac{4}{n!}\sum_{\ell=1}^{\infty}
  \left(\frac{\ell}{2\tau}\right)^{m+n-2}
  \frac{K_{m+n-2}\bigl(\ell\sqrt{1+w}/\tau\bigr)}{(1+w)^{(m+n-2)/2}}
  \right].
  \label{eq:Lnclosed}
\end{equation}
Comparing~\eqref{eq:Lnclosed} with~\eqref{eq:besselform} at parameter $m+n$ yields the
central algebraic identity of the family, valid for all $m>2$, $n\ge0$, $w>-1$ and
$\tau\ge0$:
\begin{equation}
\Lcal_n^{(m)}(w;\tau)\;=\;\binom{m+n-1}{n}\,\Lcal_0^{(m+n)}(w;\tau).
  \label{eq:crossfamily}
\end{equation}
This identity is most transparent before any of the defining integrations of
\cref{app:bessel} are performed.  Every representation of $\Lcal_0^{(m)}$ obtained there ---
the Matsubara sum~\eqref{eq:matsubara}, the Mellin integral leading to it, and the closed
form~\eqref{eq:besselform} --- descends from the single proper-time integral of
\eqref{eq:ptflow} projected onto constant fields and rescaled by $u=sk^2$,
$\hat{\mathbf p}=\mathbf p/k$:
\begin{equation}
  \kt U_k = T\sum_n\int\!\frac{\dd^3\hat p}{(2\pi)^3}\int_0^\infty\frac{\dd u}{u}\,
  \frac{u^m e^{-u}}{\Gamma(m)}\,e^{-u(\hat p^2+4\pi^2n^2\tau^2)}\;e^{-uw},
\end{equation}
in which $w$ enters only through the additive exponent $e^{-uw}$.  Proper-time
integration, spatial-momentum integration, and the Matsubara sum (or, after resummation,
the winding sum) all act \emph{linearly} on this integrand and never touch its
$w$-dependence, so the identity can be established once, at the level of the integrand
itself, before any of those integrations is carried out:
\begin{equation}
  \Lcal_0^{(m)} = \bigl[\text{linear functional}\bigr]
  \left(\frac{1}{\Gamma(m)}\,u^{m}\,e^{-u(1+\hat p^2+4\pi^2n^2\tau^2)}\;e^{-uw}\right),
\end{equation}
where ``linear functional'' denotes precisely the sequence of integrations
$T\sum_n\int d^3\hat p/(2\pi)^3\int_0^\infty du/u\,(\cdot)$ that produced
\eqref{eq:matsubara}.  Hence
\begin{equation}
  \frac{(-1)^n}{n!}\partial_w^n\!\left[\frac{u^me^{-uw}}{\Gamma(m)}\right]
  =\frac{u^{m+n}e^{-uw}}{n!\,\Gamma(m)}
  =\frac{\Gamma(m+n)}{n!\,\Gamma(m)}\cdot\frac{u^{m+n}e^{-uw}}{\Gamma(m+n)}
  =\binom{m+n-1}{n}\,\frac{u^{m+n}e^{-uw}}{\Gamma(m+n)},
\end{equation}
i.e.\ the differentiated $m$-kernel integrand is $\binom{m+n-1}{n}$ times the $(m+n)$-kernel
integrand.  Applying the linear functional to both sides gives
$\Lcal_n^{(m)}=\binom{m+n-1}{n}\Lcal_0^{(m+n)}$.  Equivalently, after all
integrations, the same statement can be verified directly on the closed forms
\eqref{eq:besselform} and~\eqref{eq:Lnclosed} using
$\Gamma(m+n)/\Gamma(m)=n!\binom{m+n-1}{n}$.  The identity has three practical consequences: (i) a single
implementation of $\Lcal_0^{(m)}(w;\tau)$ for real $m$ evaluates \emph{every}
threshold function of \emph{every} member of the family; (ii) derivative-expansion
coefficients, which involve high $n$, inherit the exponential winding convergence of
$\Lcal_0$ unchanged; (iii) structural statements (positivity, monotonicity, asymptotics)
proved for $\Lcal_0$ at all $m$ transfer verbatim to all $\Lcal_n$.

\begin{figure}[t]
\centering
\includegraphics[width=0.85\textwidth]{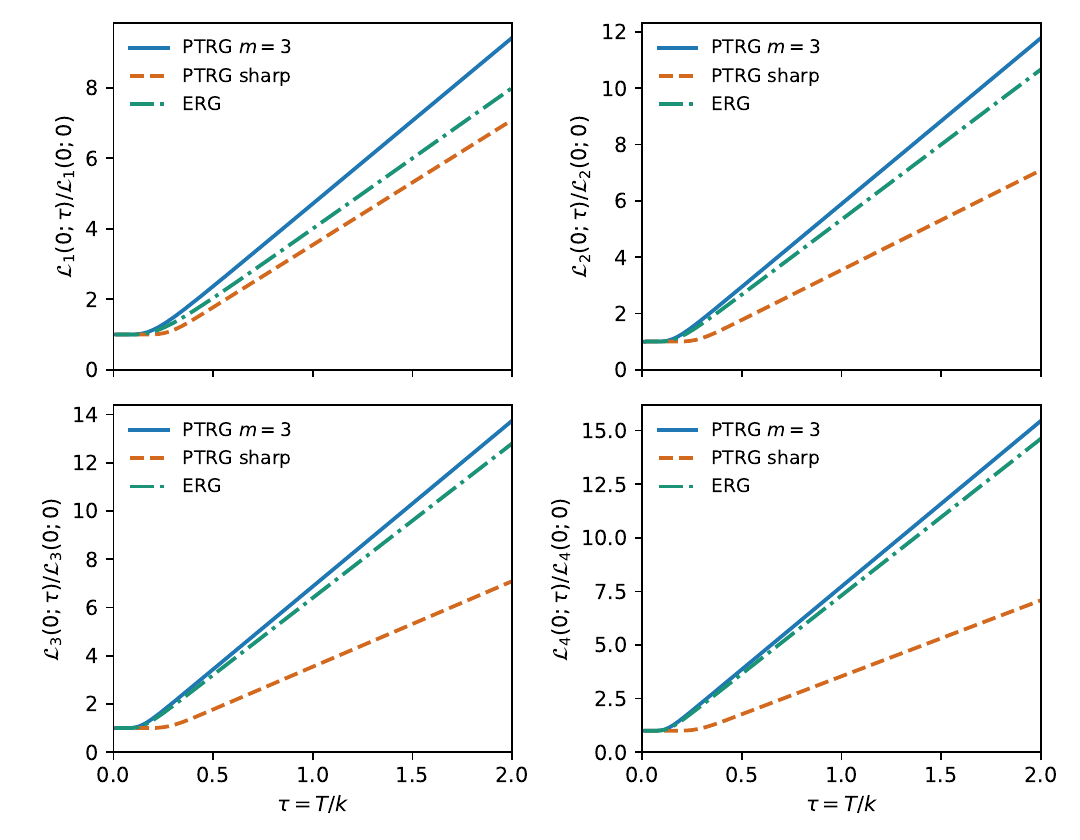}
\caption{Higher thermal threshold functions $\Lcal_n(0;\tau)$, $n=1,2,3,4$, normalised to
their $T=0$ values, $1/(32\pi^2)$ for every $n$ ($m=3$, solid blue), $1/(16\pi^2n!)$
(sharp, dashed orange) and $\Gamma(n+\tfrac12)/(12\pi^2n!\,\Gamma(\tfrac12))$
(ERG, dot-dashed green).  As in \cref{fig:L0ratio}, whose colour and
line-style key applies unchanged here.}
\label{fig:Lnratio}
\end{figure}

For the sharp regulator the situation is even simpler: its proper-time integrand is
$2\delta(u-1)e^{-uw}$ (\eqref{eq:sharpkernel} rescaled by $u=sk^2$), so each
$(-1)^n/n!\,\partial_w^n$ acting on it produces $u^n/n!\to1/n!$ under the delta function,
giving
\begin{equation}
  \Lcal_n^{\rm sharp}(w;\tau)=\frac{1}{n!}\,\Lcal_0^{\rm sharp}(w;\tau),
  \label{eq:sharpLn}
\end{equation}
i.e.\ all sharp threshold functions are proportional, with the same thermal factor
$G(\tau)$ and the same $e^{-w}$ mass dependence.

\Cref{fig:Lnratio} extends the $\Lcal_0$ comparison of \cref{fig:L0ratio} to the higher
threshold functions $\Lcal_n$, $n=1,\dots,4$.  Three structural
facts follow from the obtained results.
First, the sharp-kernel curve is \emph{exactly} the same in all four panels: by
\eqref{eq:sharpLn} the ratio $\Lcal_n^{\rm sharp}(0;\tau)/\Lcal_n^{\rm sharp}(0;0)$ equals
$\Lcal_0^{\rm sharp}(0;\tau)/\Lcal_0^{\rm sharp}(0;0)$ for every $n$, since the $n!$
cancels; the asymptotic slope stays at $2\sqrt\pi$ throughout.  Second, for $m=3$ the
$T=0$ normalisation is itself $n$-independent, $\Lcal_n^{(3)}(0;0)=1/(32\pi^2)$ for every
$n$, which is a direct consequence of~\eqref{eq:crossfamily} together with the $T=0$
limit of the closed form~\eqref{eq:besselform}, while the
dimensional-reduction coefficient in the numerator grows with $n$ (it involves
$\ell_0^{3d}$ at the shifted parameter $m+n$), so the $m=3$ slope increases with $n$, and it is given by
$2\sqrt\pi\,\Gamma(n+\tfrac32)/n!$.  Third, and in the opposite sense, the raw
high-temperature growth of $\Lcal_n^{\rm ERG}$ is itself $n$-independent,
$\Lcal_n^{\rm ERG}(0;\tau)\to\tau/(6\pi^2)$ for every $n$ (differentiating the single ERG
pole $(1+w)^{-1}$ leaves its residue unchanged), while the $T=0$ normalisation
\emph{shrinks} with $n$, $\Gamma(n+\tfrac12)/(12\pi^2n!\,\Gamma(\tfrac12))$; the ratio's
slope, $2\,n!\sqrt\pi/\Gamma(n+\tfrac12)$, therefore grows with $n$ to compensate.  The
net effect is a qualitative change of pattern relative to \cref{fig:L0ratio}: at $n=0$
the ERG was the outlier, with markedly weaker thermal response than either proper-time
member; from $n=1$ on the ERG slope climbs towards the $m=3$ one, and the \emph{sharp}
kernel becomes instead the outlier, increasingly so as $n$
grows.  In particular for $\Lcal_3$, the threshold function entering $\eta$, the exact flow and the
$m=3$ PTRG threshold have already converged closely in their thermal response, while the
sharp kernel's remains visibly softer.

\subsection{Limiting cases}
\label{sec:limits}

The closed form~\eqref{eq:besselform} makes the three physically relevant limits of the
threshold function transparent, and each provides a check against known physics.

At low temperature, $\tau\to0$, every winding term of~\eqref{eq:besselform} vanishes
as $e^{-\ell\sqrt{1+w}/\tau}$ and the threshold reduces to
\begin{equation}
  \Lcal_0^{(m)}(w;0)
  =\frac{\Gamma(m-2)}{16\pi^2\,\Gamma(m)}(1+w)^{2-m}
  =\frac{1}{16\pi^2(m-1)(m-2)}\,\frac{1}{(1+w)^{m-2}} ,
  \label{eq:T0}
\end{equation}
reproducing the known $T=0$ proper-time thresholds for this same single-term kernel
family~\citep{Bonanno:2019ukb,Litim:2002xm}.  The
approach is non-analytic in $\tau$: thermal corrections are $O(e^{-\sqrt{1+w}/\tau})$,
i.e.\ Boltzmann-suppressed with the gap of the regulated mode.  This is the correct
physics --- a massive mode at $T\ll M_{\rm eff}$ contributes only exponentially small
thermal corrections --- and it is worth noting that the smooth proper-time regulator
preserves it exactly, whereas truncated Matsubara sums do not.  This is the thermal-side
counterpart of the general observation that in compactified theories the order in which
discrete mode sums and continuous momentum integrals are performed is not innocuous: an
asymmetric treatment can hide or generate UV sensitivity~\citep{Branchina:2023rgi}.  The
closed form~\eqref{eq:besselform} performs the thermal sum exactly --- in the genuine
thermal case ($\mathbb R^3\times S^1$) the sum-to-infinity prescription is in fact
forced by the ensemble average --- so no ordering ambiguity of this kind survives
here.

In the opposite regime, $\tau\to\infty$ at fixed $w$, the $n=0$ term of
\eqref{eq:matsubara} dominates and
\begin{equation}
  \Lcal_0^{(m)}(w;\tau)
  \;=\;\tau\;\underbrace{\frac{\sqrt{\pi}\,\Gamma(m-\tfrac32)}{8\pi^2\,\Gamma(m)}\,
  (1+w)^{3/2-m}}_{\displaystyle \equiv\;\ell_0^{3d}(w)}
  \;+\;\frac{2\,\zeta(2m-3)\,\sqrt{\pi}\,\Gamma(m-\tfrac32)}{8\pi^2\,\Gamma(m)\,(2\pi)^{2m-3}}\,
  \tau^{4-2m}\;+\;O\!\left(\tau^{2-2m}\right).
  \label{eq:highT}
\end{equation}
The leading linear growth in $\tau$ is dimensional reduction: for $\tau \gg 1$ the non-zero
Matsubara modes have gap $\tau \gg 1/(2\pi)$ and the flow is that of the effective
three-dimensional theory of the zero mode, with $\ell_0^{3d}$ exactly the $T=0$
threshold function one dimension down.  For $m=3$,
$\ell_0^{3d}(w)=1/\bigl(32\pi(1+w)^{3/2}\bigr)$ and the correction term is
$\zeta(3)/(128\pi^4\tau^2)$, a coefficient we have verified numerically to $0.2\%$ at
$\tau=5$ (and improving with $\tau$).  Two features of the correction deserve comment.
The non-zero Matsubara modes decouple only \emph{algebraically} in $\tau$, as
$\tau^{4-2m}$, with a $w$-independent leading coefficient, so the corrections to
dimensional reduction are ``contact-like'' in the potential; and since the Litim ERG
threshold approaches its dimensional-reduction limit with a $\tau^{-1}$ tail (from the
$\coth$ expansion), the $m$-family at $m=3$ reduces \emph{faster} (as $\tau^{-2}$) than
the optimised ERG.

The crossover between the two regimes is set by the first non-zero Matsubara
mode, $\tau\sim1/(2\pi)$.  For
$\tau \ll 1/(2\pi)$ the mode spectrum is quasi-continuous, the sum in
\eqref{eq:matsubara} approximates the $T=0$ integral, and the threshold
stays exponentially close to its zero-temperature value; for
$\tau \gtrsim 1/(2\pi)$ the non-zero modes begin to decouple and the deviation
from unity sets in.  Because the crossover scale is fixed by the ratio
$T/k$ alone, it is common to all schemes of the family, as
seen in \cref{fig:L0ratio}: all three curves are Boltzmann-flat below
$\tau\lesssim1/(2\pi)$ and deviate together, and the scheme dependence appears
only through the power-law prefactors above the crossover.

The third limit is heavy-mode decoupling, $w\to\infty$ at fixed $\tau$: every winding
term is suppressed as $e^{-\ell\sqrt{1+w}/\tau}$ and the whole threshold decays
algebraically, $\Lcal_0^{(m)}\sim w^{2-m}$, so that modes with
$M_{\rm eff}^2\gg k^2,\,T^2$ leave the flow, in accordance with the
Appelquist--Carazzone theorem~\citep{Appelquist:1974tg}.  The sharp threshold decouples exponentially, $e^{-w}$.
This difference in decoupling rates --- algebraic versus exponential --- is the single
most important structural difference within the regulator line, and it is the origin of
most of the quantitative regulator spread seen in the applications presented 
in the next section. Taken together, the three limits show
$\Lcal_0$ interpolating between a four-dimensional quantum regime, an effectively
three-dimensional classical regime, and complete decoupling, with all crossovers
controlled analytically by the same closed expression.

\section{Consistency checks against known results}
\label{sec:flows}

With the threshold functions of \cref{sec:thresholds} available in closed form, assembling
the LPA and LPA$'$ flows is standard FRG technique.
This section states each result compactly in terms of $\Lcal_n$ and identifies where it reduces to an already established result; content that is not reducible to the existing literature in this
way is flagged explicitly as new.

\subsection{The LPA flow}
\label{sec:lpa}

The LPA retains a standard kinetic term for the field and
promotes only the potential to run, so that at constant field
$S_k^{(2)}(p)=(p^2+\omega_n^2)\mathbf{1}+\mathcal H(\phi)$, with
$\mathcal H_{ij}=\partial^2U_k/\partial\phi_i\partial\phi_j$ the field-space Hessian.  Since
the kinetic term is proportional to the identity in field space, the heat kernel factorises
over the eigenvalues $M_a^2(\phi)$ of $\mathcal H$, and the LPA flow for a generic (not
necessarily $O(N)$-symmetric) potential is a sum of single-eigenvalue thresholds,
\begin{equation}
  \kt U_k(\phi;T) \;=\; k^4\sum_{a=1}^{N}
  \Lcal_0\!\left(\frac{M_a^2(\phi)}{k^2};\,\tau\right).
  \label{eq:multifield}
\end{equation}
For the $O(N)$-invariant potential $U_k(\bar\rho)$, $\bar\rho=\phi^a\phi^a/2$, the spectrum is
one radial mode $U_k'+2\bar\rho U_k''$ and $N-1$ Goldstone modes $U_k'$, so that in the
dimensionless variables
\begin{equation}
  \rho \equiv \frac{\phi^2}{2k^2},\qquad
  u(\rho,t)\equiv\frac{U_k(\phi)}{k^4},\qquad
  \tau(t)=\frac{T}{k}=\tau_\Lambda e^{-t}
  \label{eq:conventions}
\end{equation}
where, from here on, a prime on the dimensionless potential $u$ denotes $\partial/\partial\rho$
at fixed $t$ (as opposed to the primes on the dimensionful $U_k$ above, which denote
$\partial/\partial\phi$), the flow reads
\begin{equation}
  \pt u = -4u+2\rho u'
  +\Lcal_0\bigl(u'+2\rho u'';\tau\bigr)
  +(N-1)\,\Lcal_0\bigl(u';\tau\bigr).
  \label{eq:ONflow}
\end{equation}
This is simply the usual LPA flow structure for an $O(N)$-symmetric potential: the
radial-plus-Goldstone decomposition follows from the $O(N)$ spectrum of $\mathcal H$ alone
and holds for any regulator, exact Wetterich flow included, not just for the proper-time
construction.  The only change from its familiar $T=0$ form is $T>0$ itself, entering
entirely through the threshold function $\Lcal_0$ of \cref{sec:thresholds:def} in place of
the usual zero-temperature loop integral.

One closed-form check is specific to the proper-time construction and worth recording in
full: at frozen curvature $U_k''\equiv M^2$, the normalisation~\eqref{eq:partition} forces
the flow, once integrated over all scales, to reproduce one-loop thermal perturbation theory
exactly and regulator-independently.  Working directly in the proper-time representation
before any momentum integral is performed, the winding sum of~\eqref{eq:app-afterpoisson}
splits the flow into its vacuum ($\ell=0$) and thermal ($\ell\ge1$) pieces,
$\kt U_k=\kt U_k\big|_{\ell=0}+\kt U_{k,\rm th}$; written in dimensionful proper time
$s=t/k^2$, the thermal piece alone is
\begin{equation}
  \kt U_{k,\rm th} =\frac{2}{16\pi^2\,\Gamma(m)}\sum_{\ell\ge1}
  \int_0^\infty\dd s\;s^{-3}\,\bigl(sk^2\bigr)^{m}\,
  e^{-s(k^2+M^2)}\;e^{-\ell^2/(4T^2s)} .
  \label{eq:oneloop-thermal-part}
\end{equation}
Integrating over all scales, $\int_0^\infty(\dd k/k)\,k^{2m}e^{-sk^2}=\Gamma(m)/(2s^{m})$,
removes the kernel entirely leaving
\begin{equation}
  \int_0^\infty\frac{\dd k}{k} \kt U_{k,\rm th}
  =\frac{1}{16\pi^2}\sum_{\ell\ge1}\int_0^\infty\dd s\;s^{-3}\,
  e^{-sM^2-\ell^2/(4T^2s)} ,
  \label{eq:oneloop-master}
\end{equation}
manifestly $m$-independent.  Evaluating the remaining $s$-integral with the Bessel identity~\eqref{eq:app-besselint}
gives, for every kernel of the family and for the sharp kernel
alike,
\begin{equation}
  \int_0^\infty\frac{\dd k}{k}\;
  \Bigl[k^4\,\Lcal_0\bigl(\tfrac{M^2}{k^2};\tfrac{T}{k}\bigr)
        -k^4\,\Lcal_0\bigl(\tfrac{M^2}{k^2};0\bigr)\Bigr]
  \;=\;-\,\frac{T^4}{2\pi^2}\,J_B\!\left(\frac{M^2}{T^2}\right),
  \label{eq:oneloop}
\end{equation}
where $J_B$ is the standard bosonic thermal function~\citep{Dolan:1973qd}: the closed-form
threshold reduces \emph{exactly} to the known one-loop result, with the regulator dropping
out completely, as it must.  The vacuum ($\ell=0$) piece under the same $k$-integration
reduces to $\tfrac{1}{32\pi^2}\int\dd s\,s^{-3}e^{-sM^2}$, the Schwinger proper-time
representation of $\half\Tr\ln(-\partial^2+M^2)$ per unit volume, UV-divergent at the
$s\to0$ endpoint as expected and renormalised in the standard way. 
By linearity in~\eqref{eq:multifield} the $O(N)$ generalisation is
\begin{equation}
  \int_0^\infty\frac{\dd k}{k}\Bigl[U_k(\phi)-U_{k,T=0}(\phi)\Bigr]_{\rm frozen}
  \;=\;-\frac{T^4}{2\pi^2}\Bigl[J_B\Bigl(\tfrac{M_{\rm rad}^2}{T^2}\Bigr)
  +(N-1)\,J_B\Bigl(\tfrac{M_G^2}{T^2}\Bigr)\Bigr].
  \label{eq:oneloop-ON}
\end{equation}
This simply confirms that the finite-temperature PTRG flow, once integrated over all
scales, reproduces the standard one-loop thermal effective potential exactly, as it must.

\paragraph{Quartic truncation.}
\label{sec:lpa:poly}
Projecting~\eqref{eq:ONflow} on the standard quartic ansatz gives the textbook Wilson--Fisher-type system with our thresholds in place of the usual loop integrals. In the case of a symmetric phase, $u=\bar m^2\rho+\half\bar\lambda\rho^2$,
\begin{equation}
\begin{split}
  \frac{\dd\bar m^2}{\dd t} &= -2\bar m^2-(N+2)\,\bar\lambda\,\Lcal_1(\bar m^2;\tau),\\
  \frac{\dd\bar\lambda}{\dd t} &= 2(N+8)\,\bar\lambda^2\,\Lcal_2(\bar m^2;\tau),
\end{split}
  \label{eq:sym-lpa}
\end{equation}
while in the case of a broken phase, $u=\half\bar\lambda(\rho-\kappa)^2$ with
running minimum $\kappa(t)$,
\begin{equation}
\begin{split}
  \frac{\dd\kappa}{\dd t} &= -2\kappa+3\,\Lcal_1(2\kappa\bar\lambda;\tau)+(N-1)\,\Lcal_1(0;\tau),\\
  \frac{\dd\bar\lambda}{\dd t} &= 18\bar\lambda^2\,\Lcal_2(2\kappa\bar\lambda;\tau)
  +2(N-1)\bar\lambda^2\,\Lcal_2(0;\tau),
\end{split}
  \label{eq:ssb-lpa}
\end{equation}
with the familiar $N+2$, $N+8$ combinatorial factors~\citep{Pelissetto:2000ek}.  By the
cross-family identity~\eqref{eq:crossfamily}, $\Lcal_1^{(m)}=m\,\Lcal_0^{(m+1)}$ and
$\Lcal_2^{(m)}=\tfrac12m(m+1)\,\Lcal_0^{(m+2)}$, so both systems above run entirely on
$\Lcal_0$ at shifted parameter,
\begin{equation}
\begin{split}
  \frac{\dd\bar m^2}{\dd t} &= -2\bar m^2-(N+2)\,m\,\bar\lambda\,\Lcal_0^{(m+1)}(\bar m^2;\tau),\\
  \frac{\dd\bar\lambda}{\dd t} &= (N+8)\,m(m+1)\,\bar\lambda^2\,\Lcal_0^{(m+2)}(\bar m^2;\tau),
\end{split}
  \label{eq:sym-lpa-L0}
\end{equation}
for the symmetric phase, and
\begin{equation}
\begin{split}
  \frac{\dd\kappa}{\dd t} &= -2\kappa+3m\,\Lcal_0^{(m+1)}(2\kappa\bar\lambda;\tau)
  +(N-1)\,m\,\Lcal_0^{(m+1)}(0;\tau),\\
  \frac{\dd\bar\lambda}{\dd t} &= 9m(m+1)\,\bar\lambda^2\,\Lcal_0^{(m+2)}(2\kappa\bar\lambda;\tau)
  +(N-1)\,m(m+1)\,\bar\lambda^2\,\Lcal_0^{(m+2)}(0;\tau),
\end{split}
  \label{eq:ssb-lpa-L0}
\end{equation}
for the broken phase. Note that these beta functions manifestly show a single function evaluated at two shifted regulator parameters.  Nothing here
depends on the specific threshold functions beyond their
being positive and finite: the equations, and the $N+2$/$N+8$ structure, are identical for
any regulator ever used in this context.

\paragraph{Dimensional reduction.}
\label{sec:lpa:dimred}
At high temperature, $\Lcal_n(w;\tau)\to\tau\,\ell_n^{3d}(w)$ (cf.~\cref{eq:highT}), and the
four-dimensional finite-$T$ flow reduces to the standard three-dimensional zero-temperature
LPA flow, here simply re-expressed for our thresholds:
\begin{equation}
  \pt u_3 \;=\; -3u_3+\rho_3 u_3' + \ell_0^{3d}\bigl(u_3'+2\rho_3u_3''\bigr)
  +(N-1)\,\ell_0^{3d}\bigl(u_3'\bigr),
  \qquad u_3\equiv u/\tau,\ \ \rho_3\equiv\rho/\tau,
  \label{eq:flow3d}
\end{equation}
Applying the same $\tau\to\infty$ limit termwise to~\eqref{eq:crossfamily} gives the
general-$m$, general-$n$ coefficient as a pure power law,
\begin{equation}
  \ell_n^{3d}(w) \;=\; \binom{m+n-1}{n}\,\ell_0^{3d}(w)\Big|_{m\to m+n}
  \;=\; \frac{\sqrt\pi\,\Gamma(m+n-\tfrac32)}{8\pi^2\,n!\,\Gamma(m)}\,
  (1+w)^{\frac32-m-n} ,
  \label{eq:l3dgeneral}
\end{equation}
which specialises to
\begin{equation}
  \ell_n^{3d}(w)\Big|_{m=3}=\frac{\Gamma(n+\tfrac32)}{n!\,\Gamma(\tfrac32)}\,
  \frac{1}{32\pi\,(1+w)^{n+3/2}},\qquad
  \ell_n^{3d,\rm sharp}(w)=\frac{1}{n!}\,\frac{e^{-w}}{8\pi^{3/2}} .
  \label{eq:l3d}
\end{equation}
All genuinely critical thermal physics lives in the dimensionally-reduced
flow~\eqref{eq:flow3d}, whose threshold coefficients are the general-$m$ power
laws~\eqref{eq:l3dgeneral}; for $m=3$ and for the sharp kernel these specialise to the
closed forms~\eqref{eq:l3d}.

\subsection{The LPA$'$ flow}
\label{sec:lpaprime}

Promoting the kinetic coefficient to a running, field-independent $Z_k$,
\begin{equation}
  S_k^{\rm LPA'}=\int_x\left[\frac{Z_k}{2} (\partial_\mu\phi)^2+U_k(\phi)\right]
\end{equation}
and defining the anomalous dimension $\eta\equiv-\pt\ln Z_k$, the dimensionless field is
rescaled to absorb $Z_k$ as $\rho\equiv Z_k\phi^2/(2k^2)$, so that the renormalised curvature
entering every threshold function is $w=U_k''/(Z_kk^2)$ with no further explicit
$Z_k$-dependence: every step of \cref{sec:thresholds:def} goes through unchanged with this
replacement.  The only place $\eta$ enters directly is the canonical scaling of the field
itself, which turns the
LPA canonical term $2\rho u'$ into $(2+\eta)\rho u'$.

Closing the system requires $\eta$ from the momentum dependence of the two-point function.
Expanding the proper-time heat-kernel trace to second order in a constant-background
fluctuation, and projecting the resulting one-loop self-energy onto
its $p^2$ coefficient as in the known $T=0$ construction
of~\citep{Bonanno:2019ukb}, gives, at uniform $Z_k$,
\begin{equation}
\eta(\rho;\tau)\;=\;2\rho\,\bigl(3u''+2\rho u'''\bigr)^2\,
  \Lcal_3\bigl(u'+2\rho u'';\,\tau\bigr),
  \label{eq:eta}
\end{equation}
reducing, in the quartic broken-phase truncation, to
\begin{equation}
  \eta \;=\; 18\,\kappa\bar\lambda^2\,\Lcal_3(2\kappa\bar\lambda;\tau).
  \label{eq:etaquartic}
\end{equation}
An anomalous-dimension flow of exactly this type has been derived
before, at $T=0$, with a truncated-exponential
kernel~\citep{Mazza:2001bp} and, for the same single-term kernel
family used here, in~\citep{Bonanno:2000yp,Bohr:2000gp,
Bonanno:2019ukb}.  None of these
treat the finite-temperature case in closed form, so~\eqref{eq:eta} is the genuinely new content of this
subsection: the known $T=0$ construction carried through with $\tau$ dependence
throughout, and no more.  Structurally,~\eqref{eq:eta} makes the regulator dependence of
$\eta$ transparent: for $m=3$, $\Lcal_3\sim(1+w)^{-4}$ decays algebraically in the curvature,
while for the sharp regulator it decays exponentially; this single structural difference is
the origin of most of the quantitative regulator spread reported for $\eta$ in this class of
truncations.

\paragraph{The Wilson--Fisher fixed point.}
\label{sec:fixedpoint}

\begin{figure}[t]
\centering
\includegraphics[width=0.62\textwidth]{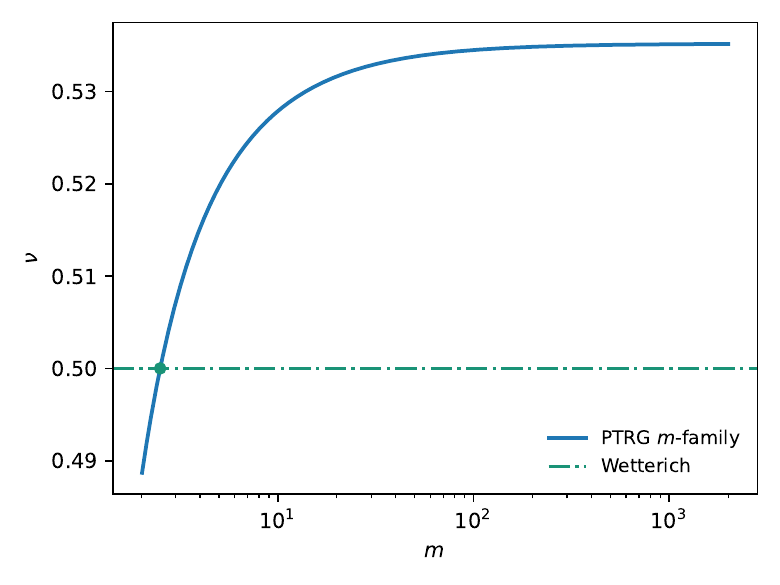}
\caption{The $\eta\equiv0$ quartic fixed point's correlation-length exponent $\nu(m)$
\eqref{eq:etazerofixedpoint}, as a continuous function of the regulator parameter $m$
(solid blue), together with the exact Wetterich flow's known closed-form value
$\nu=1/2$~\citep{Litim:2002cf} (dot-dashed green).  The two curves cross exactly at $m=5/2$
(marker), the point where the proper-time and Wetterich threshold functions coincide
identically.}
\label{fig:etazero}
\end{figure}

Substituting the high-temperature forms~\eqref{eq:l3dgeneral} and rescaling to $3d$ canonical
dimensions, $\hat\kappa=\kappa/\tau$, $\hat\lambda=\bar\lambda\tau$, turns the LPA$'$ quartic
system into the standard $d=3$ Wilson--Fisher system with our thresholds in place of the
usual ones,
\begin{equation}
  \frac{\dd\hat\kappa}{\dd t}=-(1+\eta)\hat\kappa+3\,\ell_1^{3d}(\hat w),\qquad
  \frac{\dd\hat\lambda}{\dd t}=(2\eta-1)\hat\lambda+18\hat\lambda^2\,\ell_2^{3d}(\hat w),\qquad
  \eta=18\hat\kappa\hat\lambda^2\,\ell_3^{3d}(\hat w),
  \label{eq:3dsystem}
\end{equation}
with $\hat w=2\hat\kappa\hat\lambda$.  Switching the anomalous dimension off ($\eta\equiv0$) decouples
\eqref{eq:3dsystem} into two algebraic equations, which close on a single equation for
$\hat w^*=2\hat\kappa^*\hat\lambda^*$ because $\ell_1^{3d}$ and $\ell_2^{3d}$ are both power
laws in $1+\hat w$.  This gives, for every $m>2$, a closed-form fixed point,
\begin{equation}
  \hat w^*(m) \;=\; \frac{4}{6m-7} ,
  \qquad
  \hat\kappa^*(m)=3\,\ell_1^{3d}\bigl(\hat w^*(m)\bigr),
  \qquad
  \hat\lambda^*(m)=\frac{1}{18\,\ell_2^{3d}\bigl(\hat w^*(m)\bigr)} ,
  \label{eq:etazerofixedpoint}
\end{equation}
with the correlation-length exponent defined, as usual, from the stability matrix at the
fixed point: linearising~\eqref{eq:3dsystem} (at $\eta\equiv0$) about $\hat\kappa^*(m)$,
$\hat\lambda^*(m)$ gives a single negative eigenvalue $\theta(m)$, and
$\nu(m)\equiv1/|\theta(m)|$~\citep{Litim:2002cf,Litim:2010tt}.  \Cref{fig:etazero} shows $\nu(m)$ across the
physical range, together with the corresponding exact Wetterich-flow value with the
optimised regulator, $\nu=1/2$~\citep{Litim:2002cf}.  The comparison is non-trivial rather than circular: at the single point $m=5/2$,
where~\eqref{eq:m52ergcoth} already established that our threshold function coincides
identically with the Wetterich one,~\eqref{eq:etazerofixedpoint} reproduces
$\hat\kappa^*=2/(9\pi^2)$, $\hat\lambda^*=9\pi^2/8$, and $\nu=1/2$ exactly. 

With $\eta$ switched back on,~\eqref{eq:3dsystem} still admits a closed-form
solution, for exactly the same reason as at $\eta\equiv0$: every $\ell_n^{3d}(\hat w)$ is a
pure power law, $\ell_n^{3d}(\hat w)\propto(1+\hat w)^{-p_n}$ with $p_n=m+n-\tfrac32$, cf.~
\eqref{eq:l3dgeneral}.
Writing $\hat\kappa^*=3\ell_1^{3d}(\hat w^*)/(1+\eta^*)$ and
$\hat\lambda^*=(1-2\eta^*)/(18\,\ell_2^{3d}(\hat w^*))$ from the first two equations of
\eqref{eq:3dsystem} and substituting into $\hat w^*=2\hat\kappa^*\hat\lambda^*$, the power
$p_2-p_1=1$ makes the fixed-point equation for $x^*\equiv1+\hat w^*$ linear, so $x^*$ is an explicit function
of $\eta^*$ alone.  Substituting both into the third equation,
$\eta^*=18\hat\kappa^*\hat\lambda^{*2}\ell_3^{3d}(\hat w^*)$, the resulting power of $x^*$
is $2p_2-p_1-p_3$, which vanishes identically for every $m$: the $x^*$-dependence cancels
completely, leaving a closed quadratic for $\eta^*(m)$ alone,
\begin{equation}
  (1-4B)\,\eta^{*2}+(1+4B)\,\eta^*-B=0,\qquad B(m)\equiv\frac{2m+1}{9(2m-1)},
  \label{eq:etastarquadratic}
\end{equation}
whose small positive root gives $\eta^*(m)$ exactly, after which $\hat w^*(m)$,
$\hat\kappa^*(m)$, $\hat\lambda^*(m)$ follow algebraically as above.  The
correlation-length exponent closes the same way: linearising~\eqref{eq:3dsystem} about
this fixed point, using the same power-law structure for $\partial\eta^*/\partial
\hat\kappa$ and $\partial\eta^*/\partial\hat\lambda$, gives, after the analogous
cancellations, the trace and determinant of the stability matrix in closed form,
\begin{equation}
  \mathrm{tr} = -\,\frac{2m\,\hat w^*}{1+\hat w^*} , \qquad
  \det = -\,\frac{1+4\eta^*}{1+\hat w^*} ,
  \label{eq:trdet}
\end{equation}
so that a closed form for every $m>2$ can be obtained
\begin{equation}
  \nu(m)=\frac1{|\theta_-|} , \qquad
  \theta_\pm=\frac{\mathrm{tr}\pm\sqrt{\mathrm{tr}^2-4\det}}{2}
  =\frac{-m\hat w^*\pm\sqrt{m^2\hat w^{*2}+(1+4\eta^*)(1+\hat w^*)}}{1+\hat w^*} \, .
  \label{eq:thetapm}
\end{equation}

\begin{figure}[t]
\centering
\includegraphics[width=\textwidth]{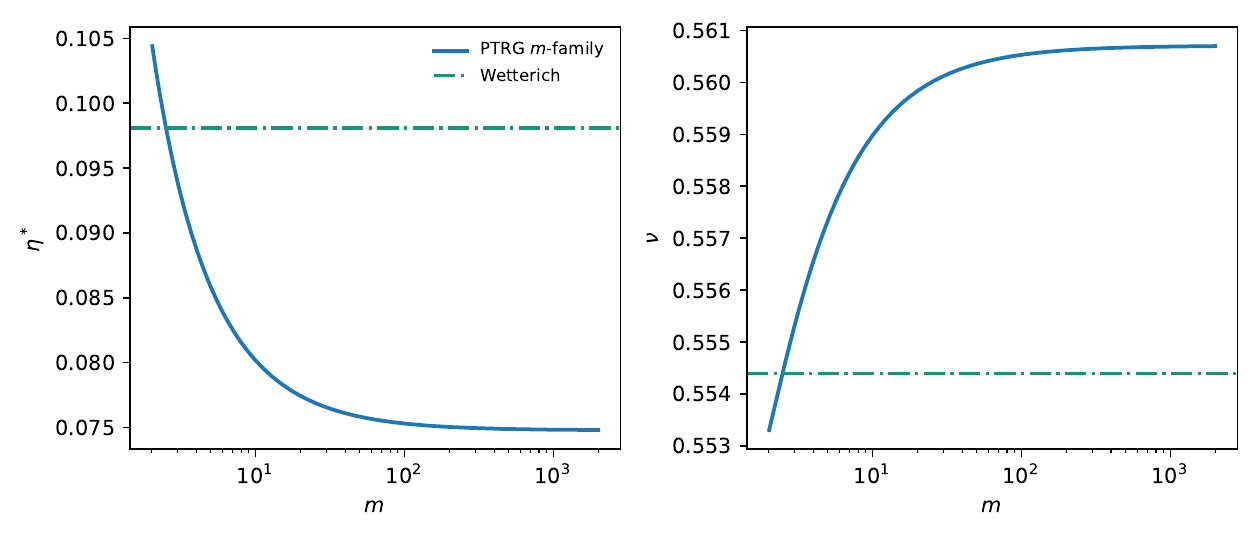}
\caption{The LPA$'$ quartic fixed point, $\eta^*$ (left) and $\nu$ (right), as a
continuous function of the regulator parameter $m$, evaluated from the closed form
\eqref{eq:etastarquadratic}--\eqref{eq:trdet} at each $m$ and plotted out to $m\sim10^3$,
where the curve has visibly converged to the sharp-kernel value.  Reference line: the exact
Wetterich flow with the optimised Litim regulator, computed from scratch here at the same
two-coupling truncation as the proper-time family.}
\label{fig:fixedpoint}
\end{figure}

\Cref{fig:fixedpoint} shows $\eta^*(m)$ and $\nu(m)$, evaluated from
\eqref{eq:etastarquadratic}--\eqref{eq:trdet}, together with the exact Wetterich
flow, computed from scratch here at the same two-coupling truncation as the proper-time
family. Continuous scans over a regulator-\emph{shape} parameter
are of course well established for the exact Wetterich equation itself, where each point
costs only an elementary momentum integral rather than a Matsubara
resummation~\citep{Canet:2002gs,Balog:2019rrg}; what is new here is
that the proper-time family reaches the same continuum of points at even lower cost, namely
solving a quadratic rather than performing any integral at
all.  The result is unambiguous: both
exponents vary smoothly and monotonically along the regulator line, with no structure,
discontinuity, or non-monotonicity anywhere between the two endpoints already discussed in
\cref{sec:thresholds:def}.  The two exponents are, however, sensitive to the regulator to very
different degrees: $\nu$ stays within $1.3\%$ of its $m=3$ value across the entire family,
while $\eta^*$ varies by close to $30\%$ from $m\to2$ to the sharp endpoint ---
consistent with $\eta^*$, unlike $\nu$, being built directly from $\Lcal_3$, whose
decoupling power varies most strongly along the line (\cref{eq:wminus1}).  Neither
exponent comes anywhere near the 3D Ising values, $\eta^*_{\rm Ising}=0.0363$ and
$\nu_{\rm Ising}=0.630$~\citep{Pelissetto:2000ek}, nor the
values~\citep{Litim:2010tt} obtain for the identical $m$-family regulator at
$O(\partial^2)$: $\nu=0.630$, $\eta=0.051$ at $m=3$ and $\nu=0.624$, $\eta=0.033$ at the
sharp endpoint.  Our own values at the same two points, $\nu\approx0.554$--$0.561$ and
$\eta^*\approx0.075$--$0.098$ (the ERG reference curve included), sit some $10$--$12\%$
low in $\nu$ and a factor of $1.8$--$2.3$ high in $\eta^*$.  The family-wide spread found
here ($1.3\%$ in $\nu$, $30\%$ in $\eta^*$) is a small fraction of either gap, so this is
not a regulator effect: it is the signature of
truncating the potential to two couplings, $\hat\kappa$ and $\hat\lambda$,
whereas~\citep{Litim:2010tt} solve for the fixed-point potential as a function of the
field with no polynomial truncation at all. 
Enriching our own truncation the same way is expected to close most of the gap for $\nu$
and, more slowly, for $\eta$, but is beyond the scope of the
present closed-form construction.

\section{Summary and discussion}
\label{sec:discussion}

We have assembled, for the $O(N)$-symmetric scalar theory with a generic potential, the
complete analytic infrastructure of the finite-temperature PTRG at LPA and
LPA$'$: the closed Bessel-$K$
winding representation of the thermal threshold functions (\cref{eq:besselform}), the
closed form and cross-family reduction of all higher thresholds
(the identity~\eqref{eq:crossfamily}), the exactly factorising sharp threshold and its identification
as the $m\to\infty$ endpoint of the family (\cref{eq:sharp}), the analytic
structure of the thresholds (complete monotonicity, the thermally enhanced convexity
singularity at $w=-1$, factorisation dichotomy, low-/high-$T$ asymptotics with explicit
subleading terms), the multi-field Hessian reduction (\cref{eq:multifield}), the anomalous
dimension (\cref{eq:eta}), and the continuous regulator-dependence of the LPA$'$ fixed point
along the entire proper-time line (\cref{fig:fixedpoint}).  Beyond assembling this infrastructure,
\cref{sec:flows} is deliberately compressed: every flow is stated in threshold-function
form and then identified with, or reduced to, an already established result rather than
re-derived or re-tabulated, and only the handful of genuinely new items above are given in
full.  Correctness rests on limiting cases rather than on any single derivation: the
$T=0$ thresholds of the proper-time literature, the exact, regulator-independent recovery
of one-loop thermal perturbation theory under frozen-curvature integration
(\cref{eq:oneloop}), the exact functional dimensional reduction (\cref{eq:flow3d}), and the
closed-form cross-check of the LPA$'$ quartic system against the known ERG fixed
point~\citep{Litim:2002cf} at $\eta\equiv0$, obtained from our own family's closed form
and found to coincide with it exactly at $m=5/2$ (\cref{fig:etazero}).

The comparison with the Wetterich equation can be summarised in one sentence per regime.
At $T=0$ and LPA the $m=3$ PTRG \emph{is} the optimised
ERG~\citep{Litim:2002xm,Bonanno:2019ukb}.  At finite temperature the
two are inequivalent schemes whose thresholds close in dual representations ($\coth$ in
frequency space versus Bessel in winding space) with different decoupling powers and
different rates of approach to dimensional reduction, except at the single point $m=5/2$
where the two coincide identically.  At LPA$'$, the $\eta\equiv0$
quartic system reduces, for the ERG, to the closed-form fixed point of~\citep{Litim:2002cf}, and, with
$\eta$ switched back on, the PTRG inherits the same crude-truncation systematics as any
other regulator, smoothly and continuously across the entire proper-time line
(\cref{fig:fixedpoint}). Distinguishing truncation error from proper-time-specific
error therefore calls for richer truncations than the one used here, not a different
regulator.

Two limitations delimit the scope of everything above.  First, the uniform-$Z_k$
LPA$'$ ignores both the field dependence of $Z_k$ and the $O(4)$-breaking split
$Z^{\rm spatial}\neq Z^{\rm temporal}$; the latter is forced at finite temperature and
untested here. Second, any quantity residing in the non-convex inner region of a
potential is coarse-graining--scheme dependent; flows applied to
first-order transitions must confine physical claims to outer-region observables or
explicitly demonstrate stability under variations of the inner-region prescription.
Within these boundaries, the formulas collected here reduce every LPA/LPA$'$ PTRG
computation at finite temperature to fast, exponentially convergent Bessel sums and a pair
of one-dimensional systems of ordinary differential equations, with each ingredient carrying an independent analytic
check.

The tools assembled here are directly applicable wherever a finite-temperature
$O(N)$-symmetric effective potential must be evolved efficiently and repeatedly.
First-order phase transitions are a natural target: nucleation rates and the
resulting gravitational-wave spectrum depend on the potential across the entire non-convex
region, and the closed-form thresholds turn what is usually an expensive numerical
Matsubara resummation into an analytic input at every point of the flow.  The same
machinery applies to the finite-temperature chiral transition in quark-meson-type models,
to condensed-matter and cold-atom realisations of the $O(N)$ universality class (superfluid
helium, Bose--Einstein condensates), and, more broadly, to any systematic assessment of
regulator-scheme dependence, since the entire proper-time line can now be scanned
continuously rather than at the handful of discrete points available before this work.

Taken together, the results above turn the finite-temperature PTRG into a
closed-form calculus: every threshold function of the entire regulator line, all
higher-order coefficients, and the complete LPA$'$ fixed-point structure are expressed
analytically.  In particular, the continuous regulator scan of
the refined truncation's fixed point --- previously accessible only at a handful of
discrete regulator values --- is obtained here in closed form for every member of the
family, providing for the first time a complete map of the regulator dependence of the
thermal PTRG observables.

\paragraph{Acknowledgements.}
The author is supported by the Estonian Research Council grants
TARISTU24-TK10, TARISTU24-TK3, and the CoE grant TK202 ``Foundations of the Universe''.

\clearpage
\appendix
\section{Derivation of the Bessel-$K$ threshold function}
\label{app:bessel}

In this appendix we derive the closed form~\eqref{eq:besselform} of the thermal threshold
function $\Lcal_0^{(m)}$, starting directly from the proper-time flow~\eqref{eq:ptflow} and
carrying the calculation through to completion.  The derivation uses only standard
textbook techniques --- a Gaussian momentum integral, a Mellin--Barnes representation, the
Jacobi/Poisson modular transformation of a theta function, and a tabulated
Bessel-function integral --- applied here, for the first time, to the general-$m$
proper-time kernel family.

Insert $S_k^{(2)}=\mathbf p^2+\omega_n^2+U_k''$ and the kernel~\eqref{eq:mkernel}
into~\eqref{eq:ptflow} at constant field:
\begin{equation}
  \kt U_k
  = T\sum_{n}\int\!\frac{\dd^3\mathbf p}{(2\pi)^3}
  \int_0^\infty\!\frac{\dd s}{s}\,\frac{(sk^2)^m e^{-sk^2}}{\Gamma(m)}\,
  e^{-s[\mathbf p^2+\omega_n^2+U_k'']} .
\end{equation}
Rescaling $u=sk^2$, $\hat{\mathbf p}=\mathbf p/k$, the proper-time integral is elementary,
$\int_0^\infty\dd u\,u^{m-1}e^{-uA_n}=\Gamma(m)A_n^{-m}$ with
$A_n=1+\hat p^2+4\pi^2n^2\tau^2+w$, and the remaining spatial integral is an Euler Beta
function,
\begin{equation}
  \int\!\frac{\dd^3\hat p}{(2\pi)^3}\,\bigl(A_n^0+\hat p^2\bigr)^{-m}
  =\frac{1}{4\pi^2}\,\frac{\Gamma(\tfrac32)\Gamma(m-\tfrac32)}{\Gamma(m)}\,
  (A_n^0)^{3/2-m}
  =\frac{\sqrt\pi\,\Gamma(m-\tfrac32)}{8\pi^2\,\Gamma(m)}\,(A_n^0)^{3/2-m},
\end{equation}
with $A_n^0=1+w+4\pi^2n^2\tau^2$.  Collecting factors turns this into the Matsubara representation~\eqref{eq:matsubara}.

To resum~\eqref{eq:matsubara} over $n$, write each Matsubara term as a Mellin--Barnes
integral,
\begin{equation}
  (A_n^0)^{3/2-m}=\frac{1}{\Gamma(m-\tfrac32)}\int_0^\infty\dd t\;
  t^{m-5/2}\,e^{-tA_n^0},
\end{equation}
valid for $m>3/2$; absolute convergence for $m>2$, $\tau>0$ justifies exchanging sum and
integral.  Under the integral, the Matsubara sum is now exactly the theta function
$\sum_ne^{-4\pi^2\tau^2tn^2}$, to which the Jacobi/Poisson inversion~\eqref{eq:jacobi}
applies, trading the sum over Matsubara modes $n$ for a sum over windings $\ell$ of the
thermal circle.  After the inversion,
\begin{equation}
  \Lcal_0^{(m)}(w;\tau)
  =\frac{1}{16\pi^2\Gamma(m)}\sum_{\ell\in\mathbb Z}
  \int_0^\infty\dd t\;t^{m-3}\,e^{-(1+w)t-\ell^2/(4\tau^2t)} ,
  \label{eq:app-afterpoisson}
\end{equation}
where the extra $t^{-1/2}$ from the inversion has shifted the Mellin power from $m-5/2$ to
$m-3$.

It remains to evaluate the winding sum in~\eqref{eq:app-afterpoisson} term by term.  The
$\ell=0$ piece is the elementary Gamma integral
$\int_0^\infty\dd t\,t^{m-3}e^{-(1+w)t}=\Gamma(m-2)\,(1+w)^{2-m}$ (finiteness requiring
$m>2$).  Every $\ell\neq0$ piece instead has the two-exponent form
$\int_0^\infty\dd t\,t^{\mu-1}e^{-at-b/t}$ (with $\mu$ here a generic Bessel order, not to
be confused with the correlation-length exponent $\nu(m)$ of \cref{sec:fixedpoint}),
tabulated as~\citep{DLMF}
\begin{equation}
  \int_0^\infty\dd t\;t^{\mu-1}e^{-at-b/t}
  =2\left(\frac ba\right)^{\mu/2}K_\mu\bigl(2\sqrt{ab}\bigr),
  \qquad a,b>0,
  \label{eq:app-besselint}
\end{equation}
with, here, $\mu=m-2$, $a=1+w$, $b=\ell^2/(4\tau^2)$, so that $2\sqrt{ab}=\ell\sqrt{1+w}/\tau$
and $(b/a)^{\mu/2}=\bigl(\ell/(2\tau\sqrt{1+w})\bigr)^{m-2}$.  Since the integral in
\eqref{eq:app-afterpoisson} depends on $\ell$ only through $\ell^2$, the windings $+\ell$
and $-\ell$ contribute identically, so summing over $\ell$ is
twice the sum over $\ell\ge1$ plus the value in $\ell=0$:
\begin{equation}
  \Lcal_0^{(m)}(w;\tau)
  =\frac{1}{16\pi^2\,\Gamma(m)}\left[
  \Gamma(m-2)\,(1+w)^{2-m}
  \;+\;4\sum_{\ell=1}^{\infty}
  \left(\frac{\ell}{2\tau\sqrt{1+w}}\right)^{m-2}
  K_{m-2}\!\left(\frac{\ell\sqrt{1+w}}{\tau}\right)\right],
\end{equation}
exactly recovering the closed form~\eqref{eq:besselform}.

\bibliographystyle{unsrt}
\bibliography{main}

\end{document}